\documentclass[apj]{emulateapj}

\shorttitle{Embedded Stellar Clusters: II}
\shortauthors{Leistra et al.}
\begin{document}

\title{The Initial Mass Functions of Four Embedded Stellar Clusters}
\author{A.~Leistra\altaffilmark{1}, A.~S.~Cotera\altaffilmark{2},
  J.~Liebert\altaffilmark{1}}
\altaffiltext{1}{Steward Observatory, University of Arizona, 933 N. Cherry 
Ave., Tucson, AZ 85721}
\altaffiltext{2}{SETI Institute, 515 N. Whisman Road, Mountain View, CA 94043}
\email{aleistra@as.arizona.edu, acotera@seti.org, jliebert@as.arizona.edu}

\begin{abstract}
We present near-infrared $J$, $H$, and $K$ images of four embedded stellar
clusters in the Galaxy.  We find a significant fraction of
pre-main-sequence stars present in at least one of the clusters.  For the
clusters dominated by main-sequence stars, we determine the initial mass
function (IMF) both by using the $K$ luminosity function and a global
extinction correction and by deriving individual extinction corrections for
each star based on their placement in the $K$ vs. $H-K$ color-magnitude
diagram.  Based on our IMFs we find a significant discrepancy between the mean
IMF derived via the different methods, suggesting that taking individual
extinctions into account is necessary to correctly derive the IMF for an
embedded cluster.

\end{abstract}


\section{INTRODUCTION}
\label{sec:intro}

Embedded clusters are increasingly recognized as vital sites of star formation
for both low- and high-mass stars.  Recent studies indicate that clusters may
account for 70-90\% of star formation and that embedded clusters (those still
partially or fully enshrouded in their natal molecular cloud) may exceed the
number of older, non-embedded open clusters by a factor of $\sim$20
\citep{elmegreen00,lada03}.  The stellar content of embedded clusters within
well-known star formation regions can now be probed, where high extinction
($A_{V} \gtrsim 10$) prohibits studies at optical wavelengths.  The IMF of such
clusters has generally been found to be consistent with a Salpeter value with a
slope of $\Gamma = -1.35$ \citep[e.g][]{okumura00,blum01,figueredo02} although
outliers have been found as well, generally with flatter slopes than the
Salpeter value \citep[e.g.][]{porras99}.

Although near-infrared (NIR) spectral classification of massive stars is
possible \citep{hanson96}, in most cases determinations of the IMF from NIR
data rely heavily, if not exclusively, on photometry and use spectroscopy only
to obtain reliable masses of the few brightest and most massive stars in a
cluster if at all.  Since these results thus depend on stellar evolutionary
models as well as details of the handling of extinction, this raises concerns
about to what extent the IMF depends on the methodology employed.
\citet{massey03} cites the example of NGC 6611, where two separate analyses of
the same data \citep{hillenbrand93,massey95} using different treatments of
extinction produced IMFs differing by more than the formal 1$\sigma$ errors
would suggest ($\Gamma = -1.1 \pm 0.1$ and $\Gamma = -0.7 \pm 0.2$.)
Similarly, the IMF for the G305+0.2 embedded cluster differs by more than the
errors between the value derived from the K luminosity function (KLF; $\Gamma =
-1.5 \pm 0.3$) and that derived using the color-magnitude diagram ($\Gamma =
-0.98 \pm 0.2$) \citep{paper1}.  Claims that variations in the IMF exist,
whether based on individual extreme clusters such as the Arches or a general
analysis of the data \citep{scalo98}, must thus be handled with care to compare
only results based on similar methodology.

The final release of the Two Micron All Sky Survey (2MASS) has fostered studies
\citep[e.g.][]{db00,db01,bdb,ivanov} which can probe a much larger portion of
the Galaxy for previously unknown embedded stellar clusters and significantly
increased the number of known embedded clusters.  A compilation of some of
these results along with previously known embedded clusters is presented in
\citet{porras03} who find that $\sim 80\%$ of the stars in their sample are
found in ``large clusters'' of more than 100 stars, despite the rarity of such
clusters.  However, these studies are not foolproof, and compilations of
``embedded clusters'' based purely on the 2MASS data without followup must be
treated with caution.  The studies based solely on stellar density criteria
\citep[e.g.][]{db00,db01} have been found in followup work by different groups
\citep[e.g.][]{db03, paper1, borissova} to have only about a 50\% success rate
toward the inner Galaxy where the stellar background is high.  We have
performed an independent search of the 2MASS archive, searching the Point
Source Catalog for regions of higher stellar density than the background
(determined locally within a 5\arcmin\ radius) which are redder in $H-K$ than
the local field.  This selects for embedded clusters, with the color criteria
helping to eliminate chance superpositions and regions of low extinction.  A
large background radius and the use of color selection are critical to the
automated identification of embedded clusters, but even color selection can
fail in regions of high background stellar density, where patchy extinction can
mimic clusters.

In \S\ref{sec:obs} we present the observations and data reduction for four
embedded clusters found in the 2MASS Point Source Catalog, in
\S\ref{sec:analysis} we present the $K$-band luminosity functions (KLF) and
initial mass functions (IMF) for the clusters, and in \S\ref{sec:compare} we
address the issue of systematic differences between different methods of
deriving the IMF for embedded clusters, for our clusters as well as IMFs from
the literature.

\section{OBSERVATIONS \& DATA REDUCTION}
\label{sec:obs}

We selected five young stellar cluster candidates from the 2MASS Point Source
Catalog based on color and density criteria as described in \citep{paper1}.  We
selected regions with a higher stellar density than the locally defined field
with redder $H-K$ color than the field to select embedded cluster candidates.
The cluster candidates were observed using the PISCES instrument \citep{pisces}
on the 6.5m MMT on Jan10-11, 2003.  PISCES uses a 1024x1024 HAWAII array with a
platescale of 0\farcs185/pixel on the MMT, providing a 3\arcmin$\times$3\arcmin\
field of view.  Images of all cluster candidates were obtained in $J$, $H$, and
$K$ filters to a limiting magnitude of $J=19.5$ , $H=18.5$, $K=18$.  Four of our
candidates (all except the one near Sh 2-217) were independently identified as
cluster candidates by \citet{bdb}, who used criteria based only on stellar
density without color considerations.  Our deeper, higher-resolution images
suggest that four of these candidates are genuine stellar clusters, while the
fifth, near Sh 2-258, contains only a few stars in the PISCES images and could
be either a very small cluster or a chance superposition.  We present our
results for the four confirmed clusters in this paper.  Seeing conditions when
the images were obtained were variable, ranging from 0\farcs5 to 1\farcs1 at
$K$.  

All images were reduced and combined using IRAF routines.  The distortions of
the PISCES camera were mapped by imaging the globular cluster NGC 4147 and
mapping the observed locations to the USNO-B known coordinates, then
constructing a transformation function using the IRAF task {\it geomap} and
correcting the images using IRAF {\it geotran}.  There are too
few USNO-B stars in the heavily extincted regions we observed to provide
suitable distortion corrections from the fields themselves.  The individual
images, taken with a spiral dither pattern, were then combined.  We fit a PSF
to each image using the IRAF task {\it psf}, allowing the PSF to vary across
the field to compensate for residual distortions.  Photometry was then done
using IRAF-DAOPHOT.

Photometric calibration was performed using the 2MASS magnitudes of field
stars.  Stars used for calibration were selected to be well-separated from
other stars and from nebulosity in the PISCES images to ensure that they were
uncontaminated in the lower-resolution 2MASS images, and to have magnitudes
bright enough to have good photometry in 2MASS $(K < 14)$.  We chose to use a
relatively large number of calibration stars rather than selecting the few most
isolated stars to reduce effects of potential variability and photometric
outliers among the calibration stars.  The scatter in the photometric
calibration derived from comparison to 2MASS is the dominant source of
photometric error, contributing two to three times the measurement errors as
reported by DAOPHOT.  DAOPHOT errors were $\sim 0.03$ magnitudes while the
calibration uncertainties were $\sim 0.1$ magnitudes.  Quoted errors in the
2MASS photometry were negligible, with most stars having an error of $\pm
0.003$ mag or less in all bands.  Thus, the quoted error should be considered
an overestimate when considering the \emph{relative} photometry of stars within
either cluster; the calibration errors from comparison to the 2MASS photometry
will shift all our measurements by the same amount.  No trend with location on
the chip was observed in the calibration for any of the clusters, though the
scatter between the PISCES and 2MASS magnitudes becomes significant in the
outermost 15~\arcsec; we thus exclude these sources from the analysis.

\section{Analysis}
\label{sec:analysis}

\subsection{Sh 2-217 Cluster}
\label{sec:217}

\begin{figure}
\includegraphics{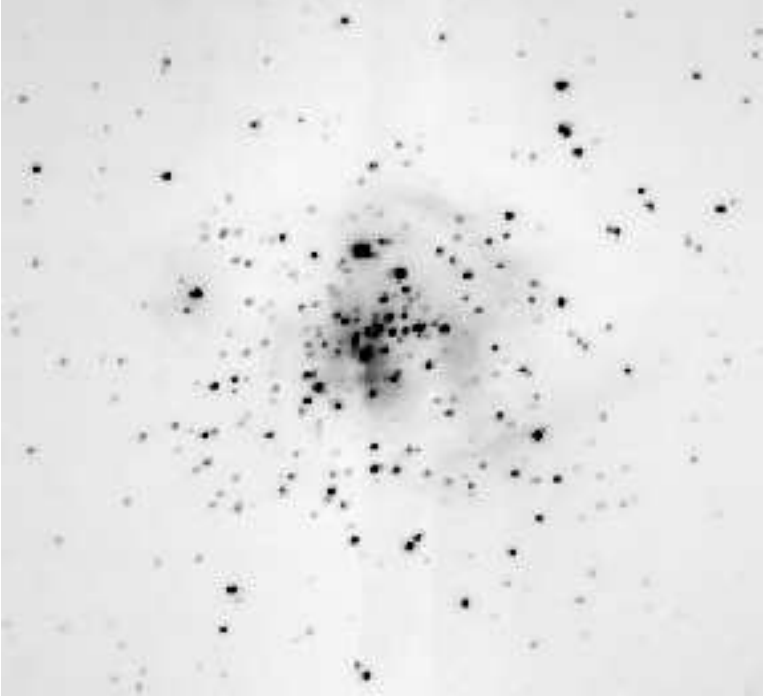} \figcaption{\label{sh217image} $K$-band image
of the Sh 2-217 cluster center (North=up, East=left).  Image is approximately
120\arcsec on a side.  The stellar density does not plateau in our entire
3\arcmin FOV, suggesting the cluster outskirts continue at least to the edge of
our image.}
\end{figure}

We present a $K$-band image of the cluster near Sh2-217 in
Figure~\ref{sh217image}.  The cluster is nearly circular in projection and is
quite dense; even in the highest-resolution individual pointings we obtained it
suffers from crowding in the central regions.  The cluster extends over most,
if not all, of the field of view. We present the $K$ image rather than a color
frame because the seeing was significantly better at $K$.

This cluster was analyzed in the NIR by \citet{deharveng}, who discuss the
large uncertainties in the distance to Sh2-217.  Based on the Lyman continuum
fluxes from the main exciting star of Sh2-217 (located several arcminutes
outside our field of view; the cluster is located on the periphery of the H II
region) they adopt a distance of $5.0 \pm 0.8$ kpc., which is consistent with
the kinematic distance to the associated molecular gas.  The cluster is
coincident with a peak in the 8~$\micron\ $ emission as measured by the MSX
mission, suggesting dust is still present in the cluster.

\begin{figure}
\scalebox{0.75}{\includegraphics[angle=-90]{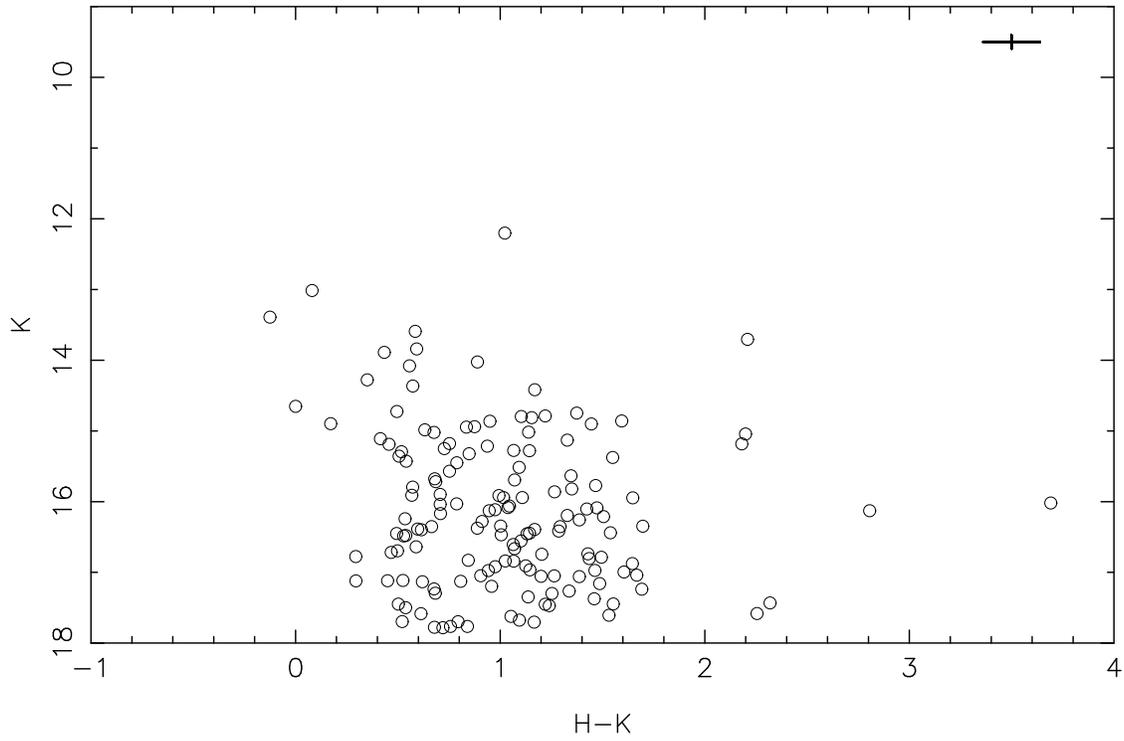}} \figcaption{\label{217cmd}
$K$ vs. $H-K$ color-magnitude diagram for the cluster near Sh2-217. Average
DAOPHOT error bar is the size of the plot symbols or smaller.  Overall
uncertainties including calibration (which include terms that will not affect
the relative position of the points) are indicated by the symbol in the upper right.}

\end{figure}

\begin{figure}
\scalebox{0.75}{\includegraphics[angle=-90]{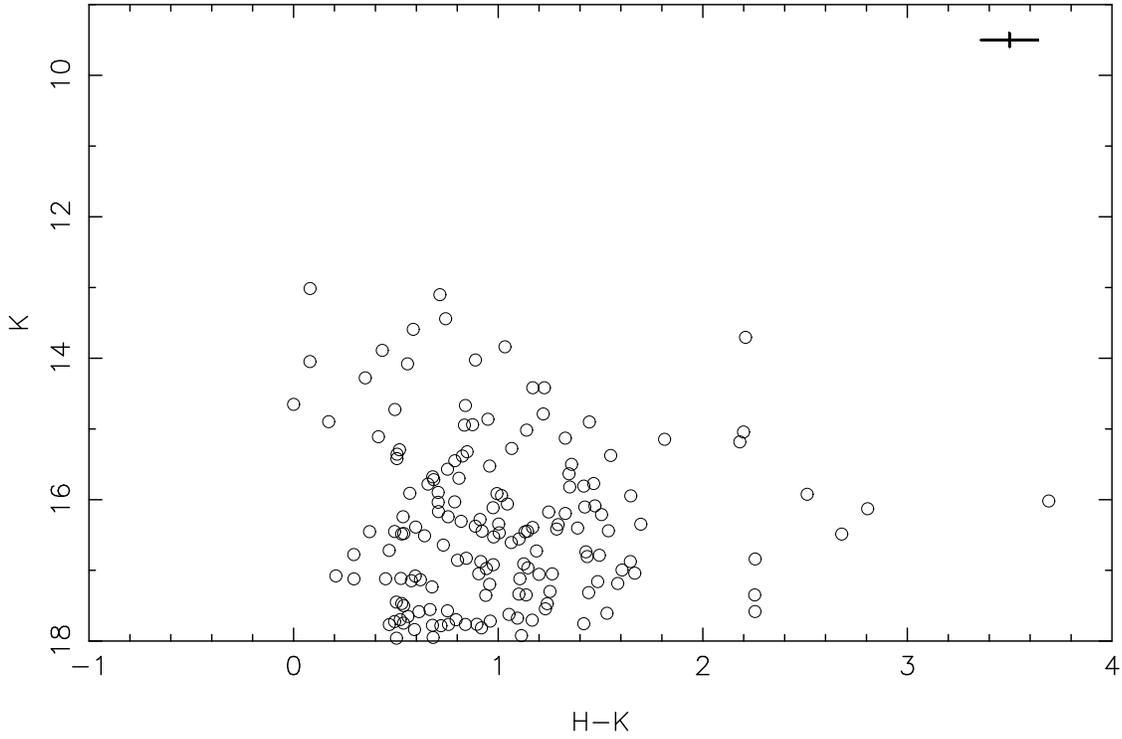}}
\figcaption{\label{217-statcorr} Statistically-corrected $K$ vs. $H-K$
color-magnitude diagram for the cluster near Sh2-217.  Statistical correction
was done based on the 2MASS Point Source Catalog for an adjacent field,
extrapolating to the our limiting magnitude based on the luminosity function.} 
\end{figure}

The $K$ vs $H-K$ color-magnitude diagram is shown in Figure~\ref{217cmd}.  The
stellar density (determined in $K$ where the seeing was best) does not plateau
in the $3\arcmin\times3$\arcmin\ field of view, suggesting that the true
field-star density level has not been reached, and cluster stars are still
present out to the edges of the field.  As a result, in order to correct for
foreground contamination, we selected an adjacent field of the same size from
2MASS to use as a comparison field.  We assumed the luminosity function of the
field to be the same as that of the outer portions of the cluster (excluding
the inner regions to minimize the results of crowding and mass segregation) in
order to extrapolate from the limiting magnitude of 2MASS to that of our
images.  We then binned the field stars by $K$ and $H-K$ with a bin size of 0.5
magnitudes and randomly selected the appropriate number of stars for removal
from each bin in the cluster region.  This is similar to the procedure employed
by, among others, \citep{blum00} (who do not describe an extinction correction;
this lack of correction is equivalent to assuming a common extinction) and
\citet{figueredo02}, and is the method we employed in \citep{paper1}.  The
resulting statistically corrected CMD is shown in Figure~\ref{217-statcorr}.  A
total of 62 stars were removed in this procedure, out of an initial total of
236.  The fairly wide distribution in $H-K$ for cluster stars is likely due to
a combination of factors, notably an actual spread due to differential
extinction to different regions of the cluster and to the greater influence of
crowding in $H$ (where the seeing was poorer).  Although we find that
individually correcting extinctions generally provides a superior estimation of
the IMF compared with using only the $K$ data and a single extinction for the
cluster as a whole, in this situation the lower quality (in particular the
poorer seeing and consequently more severe crowding) of the $H$-band data leads
us expect that the $K$ luminosity function (KLF) will produce a better estimate
of the IMF for this cluster than the CMD will.  We have previously compared
these two methods of determining the IMF for embedded clusters in
\citet{paper1}; in that case, we found they gave different results, with the
``CMD'' method (which we anticipate will be more reliable in most cases,
especially where variable extinction is present) yielding a flatter slope.

\subsubsection{The KLF}

In order to obtain a robust determination of the KLF and the IMF, we need to
determine the completeness of our data.  To do this, we performed artificial
star tests.  We inserted five artificial stars, each of the same magnitude, at
a time into the cluster region, then ran IRAF-DAOPHOT with the same parameters
as we used for the initial analysis.  This procedure was repeated 50 times for
each magnitude bin ($\Delta m = 0.5$), for a total of 250 artificial stars
added in each bin in $H$ and in $K$.  The stars were added in small numbers at
a time to avoid having the artificial stars significantly change the crowding
characteristics and thus influence the completeness.  The artificial star tests
indicate a high level of completeness down to $K=17.5$.  The actual
completeness is most likely slightly lower, since in the crowded central region
of the cluster our method may produce false positives when the artificial star
is placed on top of a real star of approximately the same magnitude.  Despite
this concern we have used the calculated incompleteness in correcting the KLF;
however, we have excluded the $K=17.5$ bin from consideration, both because
this effect will be most pronounced at faint magnitudes, and because
statistical uncertainties in the incompleteness will be significant.

\begin{figure}
\scalebox{0.75}{\includegraphics[angle=-90]{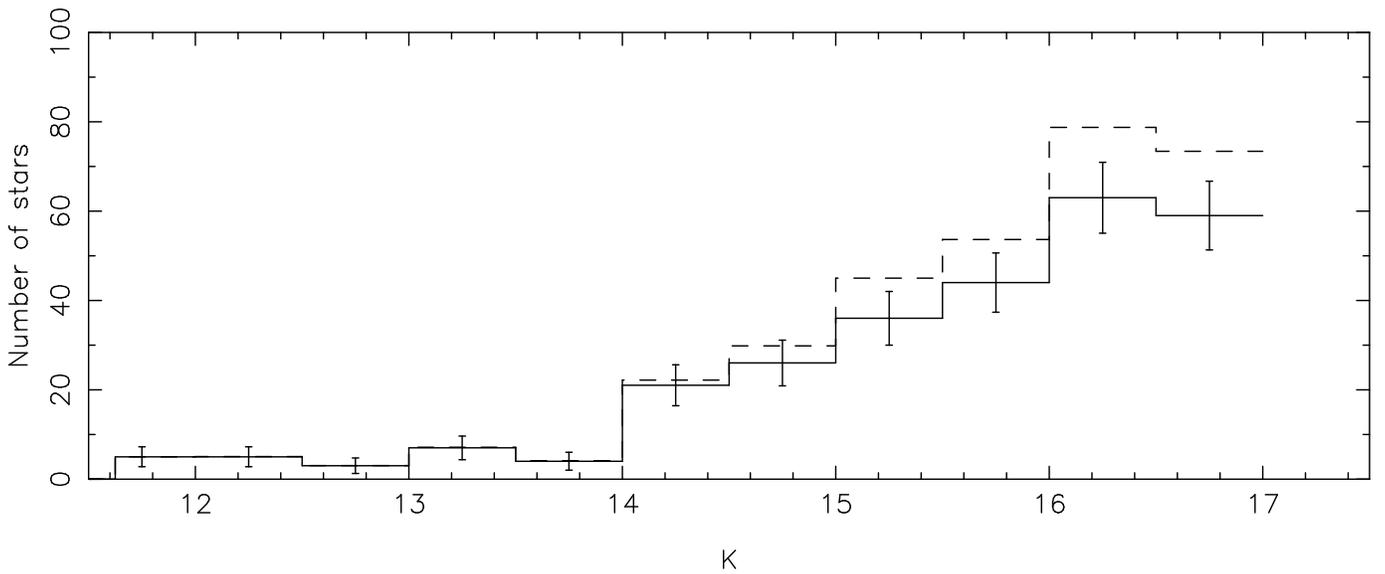}}
\figcaption{\label{sh217-KLF} $K$ luminosity function for the Sh2-217 cluster.
(Solid is measured; dashed is corrected for incompleteness.)  Since the cluster
fills the field of view and the comparison 2MASS data do not go as deep, we do
not show a field sample for comparison.}
\end{figure}

Knowing our incompleteness, we can calculate the KLF for the cluster.
Figure~\ref{sh217-KLF} shows both the uncorrected and completeness-corrected
versions of the KLF, as derived from all sources detected in $K$.  The slope of
the KLF is $0.31 \pm 0.04$, with no extinction correction applied.

\subsubsection{The IMF}

We derived an IMF for the Sh 2-217 cluster by two methods.  For both methods we
used a distance to the cluster of 5 kpc \citep{deharveng}.  In general we
believe that the ``CMD method'' for deriving the IMF, which uses
individually-derived extinctions for each star in the cluster, to be more
reliable than the ``KLF method'' which assumes a common extinction to all stars
in the cluster, since variable extinction is frequently apparent in the NIR
images of embedded clusters.  However, in this case our $K$ data is superior to
our $H$ data due to the difference in the seeing, which was $\sim 0.\arcsec$
in $K$ and $\sim 1.2\arcsec$ in $H$.  This suggests that despite the general
drawbacks of the KLF method it may be preferable in this situation; including
the $H$-band data adds nothing if it is of poor quality.  At the least, using
both methods will provide information on potential systematic effects in the
IMF determination that depend on the methodology used.

The KLF method of determining the IMF is sometimes employed even when
multi-color photometry and spectra are available \citep[e.g.][]{blum00}, and is
simply a transformation from $K$ magnitude bins to mass bins.  The major
problem with this is that of variable extinction, which for many embedded
clusters is significant even in $K$.  To make this transformation, we first
correct the observed $K$ for distance and extinction.  Without spectra, we
cannot obtain a precise estimate for the extinction; instead, we compute an
average extinction correction based on the observed $J-H$ and $H-K$ colors of
the brighter stars.  Since there is little difference in the intrinsic $H-K$
color of stars of spectral type F5 and earlier, this estimate of an average
extinction is not sensitive to minor errors in the distance estimate.  Using
the stellar evolutionary models of \citet{newmodels} for solar metallicity, we
relate the mass for each track to an absolute $K$ magnitude for a star on the
ZAMS.  We transformed $L_{bol}$ to $K$ using the bolometric corrections from
\citet{vacca} for the early spectral types and \citet{malagnini} for later
spectral types.  We then use the intrinsic $V-K$ colors from \citet{bbrett} for
A-M stars and from \citet{wegner} for O and B stars.  Finally we interpolate
linearly between the masses available on the evolutionary tracks to find the
masses corresponding to our magnitude bins, and fit a power law to the
resulting mass function.  The IMF slope we derive by this method is $\Gamma =
-2.7 \pm 0.25$, excluding bins corresponding to $K > 17.5$ where incompleteness
becomes significant.  The slope we fit to the KLF itself for sources detected
in both $H$ and $K$ is $0.35 \pm 0.04$.

\begin{figure}
\scalebox{0.75}{\includegraphics[angle=-90]{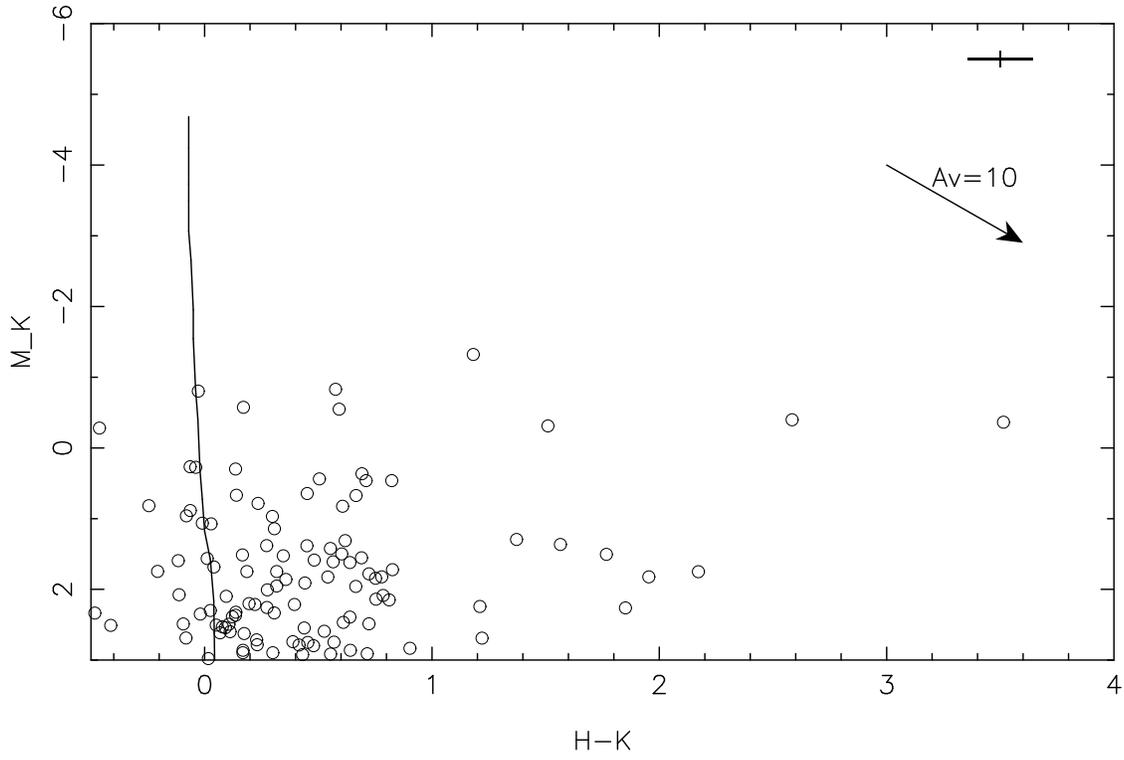}}
\figcaption{\label{ZAMSa} $K$ vs. $H-K$ color-magnitude diagram for
the Sh2-217 cluster adjusted to a distance of $5$ kpc.  The ZAMS converted to
observed quantities as described is overplotted.}
\end{figure}

We have previously used multi-color photometry in conjunction with near-IR
spectroscopy of the brightest stars to determine the IMF for embedded clusters
\citep{paper1}.  Although we do not have spectra in this case, we can still
derive extinctions for individual stars based on their near-IR colors.  We use
the same evolutionary models and conversions from theoretical to observed
quantities described for the KLF method.  The ZAMS derived from the
evolutionary tracks of \citet{newmodels} is overplotted on the distance and
extinction-corrected CMD in Figure~\ref{ZAMSa}.  The ZAMS lies in the middle of
the distribution of stars due to the method used to estimate an average
extinction.    The scatter around the ZAMS is rather large, and is likely due
to a combination of variable extinction in the cluster region and poor
photometry in $H$, especially in the central portion of the cluster.

Since extinction appears to vary across the cluster, we impose an extinction
limit on the sample used for the CMD computation of the IMF.  Since the most
massive stars can be seen to greater extinction than less massive stars,
neglecting to impose this constraint will produce an overly flat IMF.  Thus we
need to simultaneously limit our sample by extinction and by mass.  We use a
mean extinction to the Sh2-217 cluster of $A_{V} = 9$ mag as determined by
individually de-reddening sources until they reach the main sequence.  With the
exception of a few extreme outliers that most likely suffer from poor
photometry, most sources with higher extinctions have $A_{V} < 15$.   At a
distance of 5 kpc with our limiting magnitudes, we can observe stars earlier
than G4 to an extinction of 9 mag and earlier than F0 to an extinction of 15 mag.

At a distance of 5 kpc with our limiting magnitude, an extinction of $A_V = 9$
limits us to G4 and earlier stars, while $A_V = 15$ limits us to F0 and
earlier.  We select $A_V = 10$ and G2 as our limits; stars with higher
extinction or later spectral type cannot be detected over the entire range of
mass or extinction included and thus are excluded.  Approximately 34 stars have
a higher extinction than this, including five with extreme calculated
extinctions ($A_V > 50$) that most likely suffer from poor photometry and have
unrealistic colors.  This extinction limit will also have the effect of
excluding stars with $K$-band excess from the IMF determination, since such
sources would appear to be at high extinction.  This will tend to push the IMF
to flatter values, since lower-mass sources spend more time as IR-excess
objects and thus are more likely to be ruled out by this criterion.  However,
we consider this to be a better approach than including the sources since: 1.
the number of sources detected in all three bands in this cluster showing
near-IR excess is small, suggesting such sources will not significantly
influence the mass; 2.  the majority are of a low enough mass to fall below the
completeness limit, and are thus excluded anyway regardless of the method; 3.
including them (since not all sources are detected in $J$) would weaken the
extinction limit and tend to again force the IMF to flatter values.  In
clusters where IR-excess sources dominate we do not fit an IMF (see Section
\ref{sec:pms} for further discussion).

Individual extinctions are derived by moving the stars
along the direction of the reddening vector until they lie on the ZAMS.  Once
the stars have been corrected individually for extinction, they are placed in
mass bins.  We fit a power law to the data, excluding masses $< 1.1 M_{\odot}$
from the fit since they cannot be seen over the entire range of extinctions in
the cluster.  The IMF slope we derive by this method is $\Gamma = -1.61 \pm
0.2$.  As for the KLF method, the quoted errors represent only the formal
errors in the fit and should be considered an underestimate.

The difference between these two values for the IMF emphasizes that formal
statistical errors significantly underestimate the true uncertainties in the
IMF.  In this case, as was the case in \citet{paper1}, the CMD method gives a
noticeably flatter result than the KLF method.  Clearly the individual
extinction correction leads to the conclusion that more massive stars are
present than an average correction does.  This could be due to the effects of
mass segregation, or to an incorrectly chosen extinction limit (so that we
truly are sampling massive stars more completely than lower mass stars).  It is
difficult to understand which of these effects is most important without
obtaining spectra for a significant number of stars in the cluster. 

This cluster is also analyzed by \citet{porras99}, who use a slightly different
distance ($5.8$ kpc) and extinction ($<A_{V}> = 5.3 \pm 3.7$) and derive an IMF
slope of $\Gamma = -0.59$ based on 54 sources using the $J$ vs $J-H$ CMD to
individually correct extinctions and compare with a theoretical JLF.  This is a
significant discrepancy from our result with either method.  A number of
factors may contribute to this difference.  The most significant, however, is
likely to be due to a different choice of cluster boundaries.  Their quoted
cluster radius corresponds to only 50\farcs9, compared to ours of $\sim
80\arcsec$.  This suggests that their IMF will be more weighted toward the
cluster core than ours.  The value they quote for a field + cluster IMF is
$\Gamma = -2.71 \pm 0.24$, much steeper and in fact quite close to the value we
obtain using the KLF.  If the cluster suffers from mass segregation, as we
would expect given that it is observed even in very young clusters (e.g. the
Arches cluster \citep{stolte02}), we expect the core IMF to be quite flat.
When we re-derive the IMF for this cluster using their radius with our data, we
derive an IMF of $\Gamma = -1.55 \pm 0.22$, statistically indistinguishable
from our original result.  \citet{porras99} do not comment on issues of
confusion or field star contamination, so we cannot evaluate how much of an
effect it is likely to have on their result; we expect crowding to be a more
significant issue, and a misidentification of blended sources as single stars
by \citet{porras99} could account for their finding a steeper IMF than we do
using the same method.

\subsection{IRAS 06058+2158 Cluster}

\begin{figure}
\includegraphics{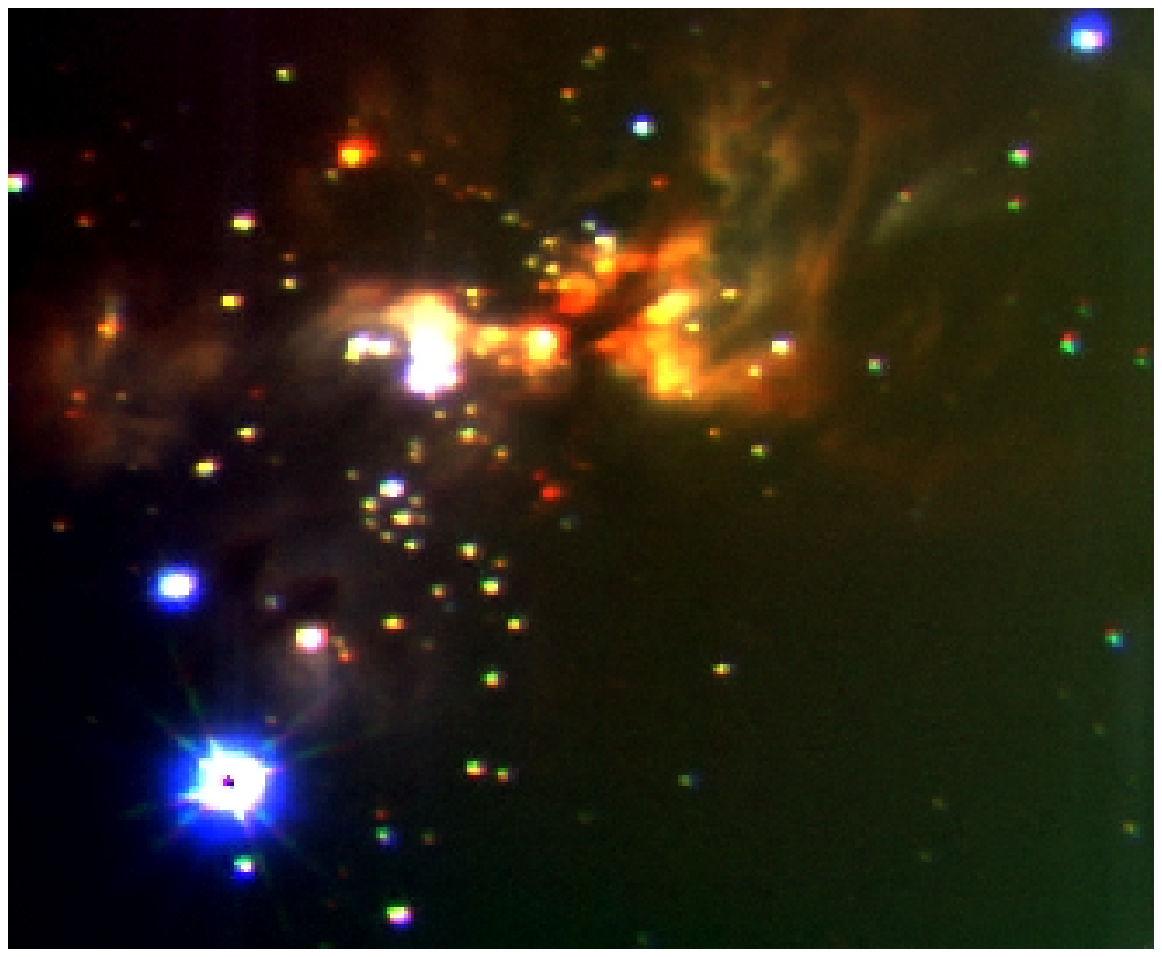}
\figcaption{\label{6058image}  Color composite ($J$ = blue, $H$ = green, $K$ =
red) of the IRAS 06058+21 cluster region (North=up, East=left).  Image is approximately 140\arcsec on a
side.}
\end{figure}

We present a three-color composite of the cluster near the IRAS source
06058+2138 in Figure~\ref{6058image}.  \citet{bik05} obtained VLT spectra in
$K$ of several NIR point sources near the IRAS point source, which is located
near the center of the cluster, and identified two candidate massive YSOs and
an embedded early-B star.  The spectrophotometric distance they derive from the
B star is 1.0-1.5 kpc.  This cluster is much more heavily embedded than
the Sh 2-217 cluster, with significant nebular emission and prominent dust lanes.
Numerous OH and methanol masers have been detected in this region, which along
with the IRAS point source suggest ongoing star formation. (see, e.g.,
\citet{caswell}, \citet{shk2000}).  A peak in the 8~$\micron\ $ emission is
observed in the MSX data, extending from the region of NIR nebulosity to the
southeast to the isolated bright source.
 
The embedded cluster here is described in the compilation of \citet{lada03},
who quote a distance of $1.5$ kpc \citep{carpenter93}.  However, \citet{hansonuchii} describe a
UCHII region associated with the IRAS point source, and quote a distance of
$2.2$ kpc \citep{koempe}, and \citet{bik05} obtain a spectroscopic distance of 1.0-1.5 kpc.
With no {\it a priori} reason to prefer one distance over the other, and no
uncertainties associated with either, we use the average distance of $1.5$ kpc
for our analysis.

\subsubsection{The KLF}
\label{sec:pms}

\begin{figure}
\scalebox{0.75}{\includegraphics[angle=-90]{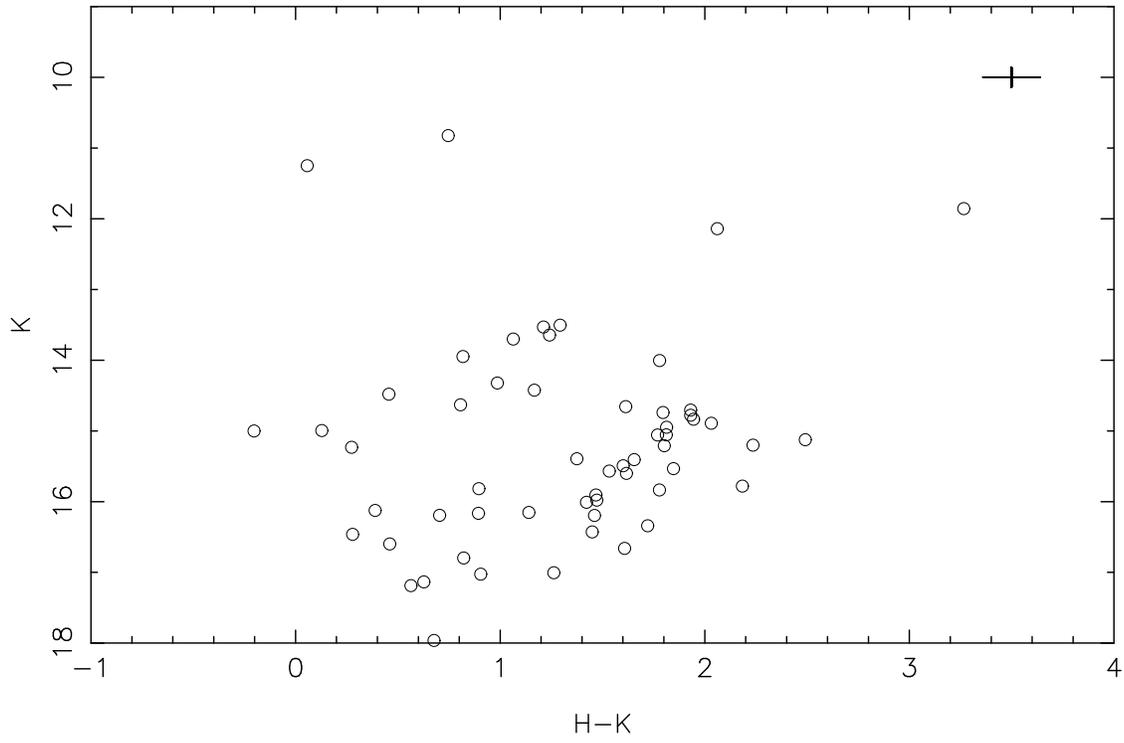}}
\figcaption{\label{6058cmd}Statistically field-star corrected $K$ vs. $H-K$ color-magnitude diagram for
the cluster near IRAS 06058+2158.   Significant and variable extinction is
evident in this cluster. Typical errorbars are the size of the plot symbols. Overall
uncertainties including calibration (which include terms that will not affect
the relative position of the points) are indicated by the symbol in the upper right.}
\end{figure}

\begin{figure}
\scalebox{0.75}{\includegraphics[angle=-90]{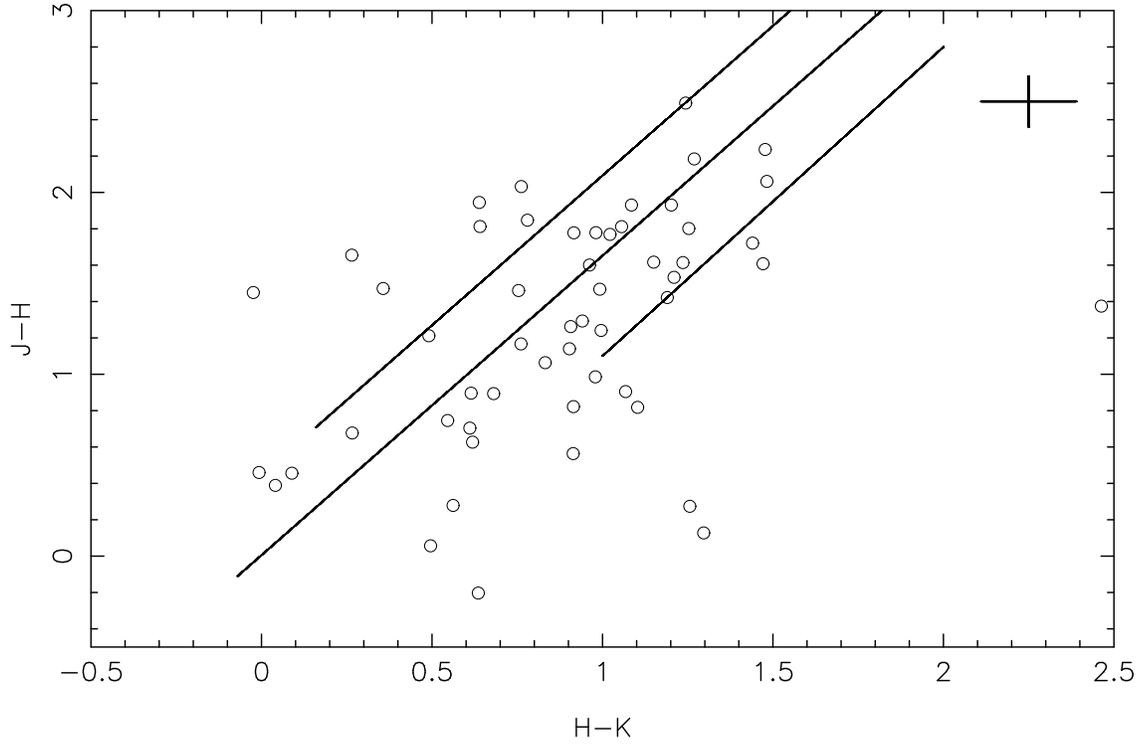}}
\figcaption{\label{6058cc} $J-H$ vs. $H-K$ color-color diagram for the cluster
near IRAS 06058+2158.  The lines (parallel to the reddening vector) delineate
the possible location of reddened main-sequence stars.  Over half of our
sources show $K$-band excess; their colors are not consistent with reddened
main-sequence sources.  Typical errorbars are the size of the plot symbols. Overall
uncertainties including calibration (which include terms that will not affect
the relative position of the points) are indicated by the symbol in the upper right.
Points to the left of the reddening lines lie in crowded regions and may suffer
from confused photometry.  Points more than 3 $\sigma$ to the left of the
reddening line were excluded from the IMF determination.}
\end{figure}

We present color-color and color-magnitude diagrams for this cluster in
Figures~\ref{6058cmd} and \ref{6058cc}.  The ``cluster region'' was defined to
coincide with the extent of the near-IR nebulosity, and field stars were
statistically corrected as described in \ref{sec:217}.  The CMD shows objects
spanning a range of extinctions, with relatively few objects with colors
consistent with unextincted main-sequence stars remaining in the statistically
corrected data.  Since the cluster in this case did not fill the field of view,
we were able to use the data from the non-cluster portions of the region as our
field, eliminating the need for an off-source 2MASS field and extrapolation
based on the KLF.  The $J-H$ vs. $H-K$ color-color diagram shows that the
majority of sources in the cluster region exhibit $K$-band excess and fall in
the region populated by reddened CTTS and YSOs, suggesting they may be
pre-main-sequence objects.  Because the amount of $K$-band excess is affected
by many factors (see, e.g.,\citet{meyer97}), deriving masses for these objects
is difficult.  We thus derive only a KLF for this cluster, and do not convert
it to an IMF or derive an IMF from the CMD.  We determine and correct for our
incompleteness as in \ref{sec:217}.  The KLF we derive using a
statistically-corrected sample of all sources detected in $K$, with no attempt
to correct for extinction due to the uncertainty of the intrinsic $H-K$ color
of the YSOs that are present, has a slope of $0.30 \pm 0.03$.  If only sources
detected in $H$ are included, the KLF declines for $K > 14.5$ since the high
extinction means the fainter objects are less likely to be detected in $H$.

\subsubsection{Pre-Main-Sequence Objects}

A total of 37 out of 58 sources (63\%) detected in all three bands show a
K-band excess in the color-color diagram, suggesting they are pre-main sequence
objects.  This is a lower limit on the actual pre-main-sequence fraction of the
cluster, since objects with a sufficiently high IR excess may be detected in
$K$ but not in $J$ or $H$.  A total of 49 sources were detected in $K$ within
the cluster region that were undetected in $J$, $H$, or both.  Adding in these
sources would give a PMS fraction of 80\%.  The latter figure is an upper limit, since
some of the $K-$-only detections are likely to be knots of nebular Br $\gamma$
emission or heavily extincted background stars.   Comparing these values to the
near-IR excess fraction of embedded clusters of known ages presented in
\citet{haisch2001}, we conclude that the age of the IRAS 6058 cluster is less
than 3 Myr. 

We observe a few sources with colors even redder than the reddened extension of
the CTTS locus.  \citet{meyer97} observe sources with similar colors, and
suggest re-radiation by an extended envelope as an explanation.

\subsection{IRAS 06104+1524 Cluster}
\label{sec:6104analysis}

\begin{figure}
\includegraphics{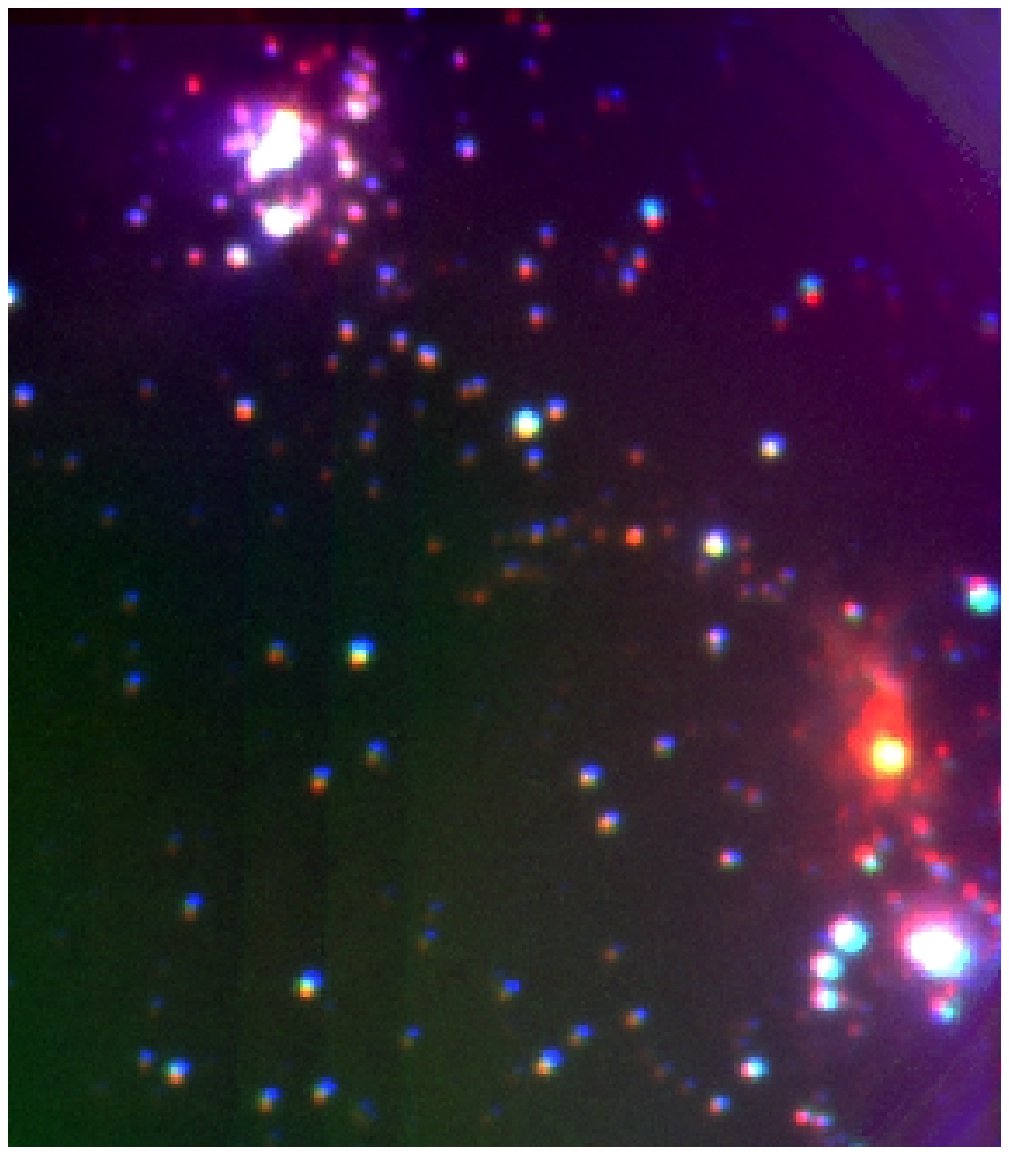}
\figcaption{\label{6104image} Color composite ($J$ = blue, $H$ = green, $K$ =
red) of the IRAS 06103+1523 / 06104+1524 cluster(s) (North=up, East=left).
Image is approximately 120\arcsec on a side.  }
\end{figure}

\begin{figure}
\scalebox{0.75}{\includegraphics[angle=-90]{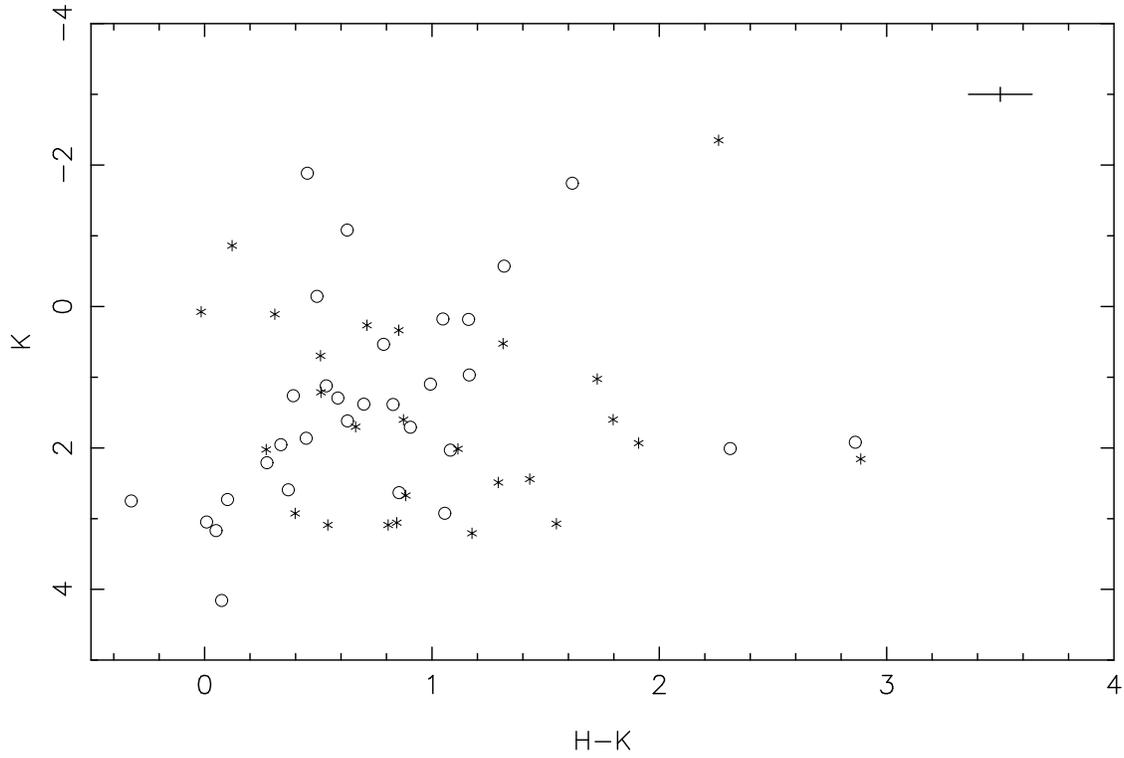}} 
\figcaption{\label{6104cmd} $K$ vs. $H-K$ color-magnitude diagram for
the IRAS 06104+1524 / 06103+1523 cluster(s) adjusted to a distance of $3.5$
kpc.  {\it Circles}: NE cluster.  {\it Asterisks}: SW cluster.  There is no
difference in the CMD apparent for the two clusters, so we treat them as one to
improve the statistics. Typical
errorbars are the size of the plot symbols. Overall
uncertainties including calibration (which include terms that will not affect
the relative position of the points) are indicated by the symbol in the upper right.}
\end{figure}

\begin{figure}
\scalebox{0.75}{\includegraphics[angle=-90]{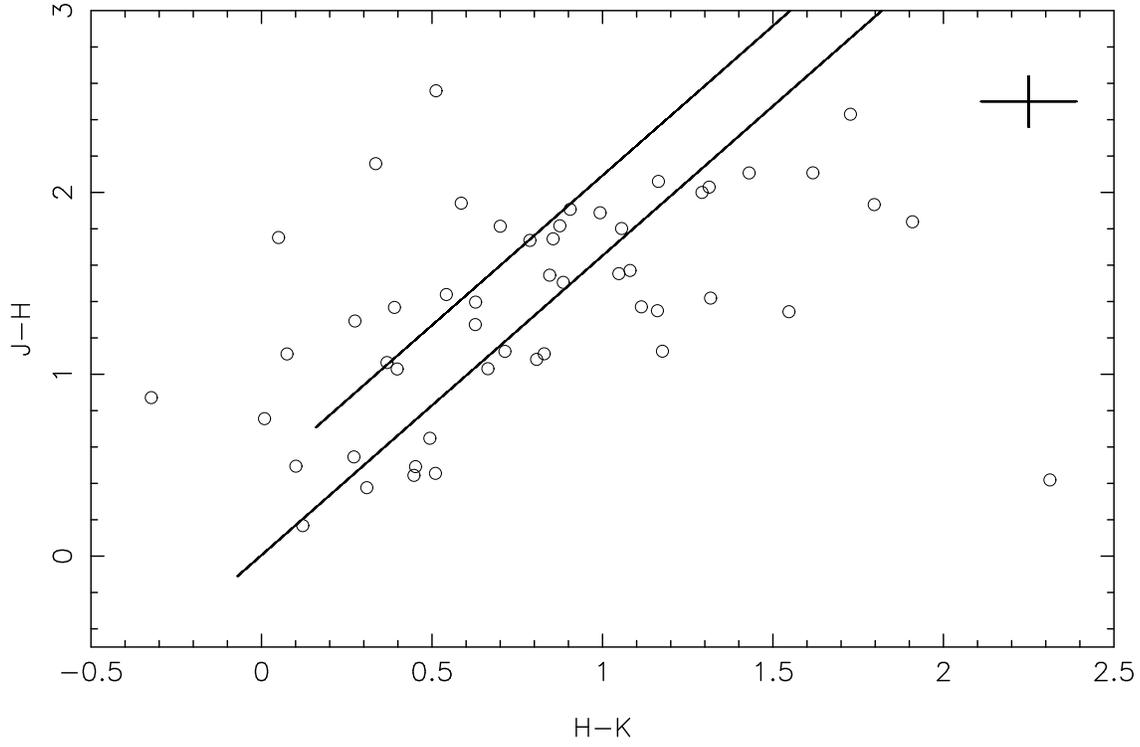}} 
\figcaption{\label{6104cc} $J-H$ vs. $H-K$
color-color diagram for the IRAS 06104+1524 / 06103+1523 cluster(s). Typical
errorbars are the size of the plot symbols. Overall
uncertainties including calibration (which include terms that will not affect
the relative position of the points) are indicated by the symbol in the upper right.}
\end{figure}

\begin{figure}
\scalebox{0.75}{\includegraphics[angle=-90]{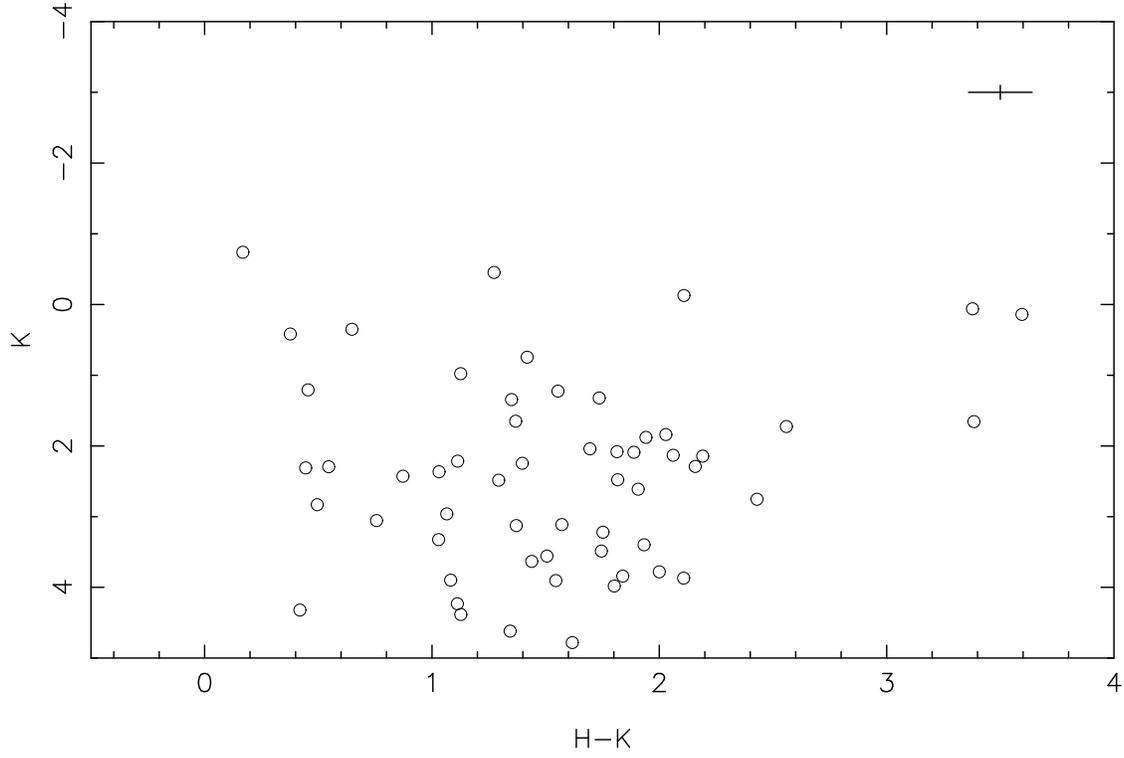}}
\figcaption{\label{6104-statcorr} Statistically corrected  $K$ vs. $H-K$ color-magnitude diagram for
the IRAS 06104+1524 / 06103+1523 cluster(s) adjusted to a distance of $3.5$
kpc.  Typical
errorbars are the size of the plot symbols.Overall
uncertainties including calibration (which include terms that will not affect
the relative position of the points) are indicated by the symbol in the upper right.} 
\end{figure}

The near-IR cluster image (Figure~\ref{6104image}) of the cluster near IRAS
06104+1524 shows a clear separation into two subclusters separated by $\sim
2\arcmin$.  The southwest subcluster is dominated by two closely spaced bright
sources while the northeast subcluster is denser and is not dominated by a
single object.  A ridge of marginally higher density than the surrounding field
appears to lie between the subclusters, though it is not apparent whether this
is a real feature.  The MSX 8~$\micron\ $image similarly shows two separate
peaks, with no indication of a connection.  These are treated as separate
clusters by \citet{bdb}, and there are IRAS point sources associated with each
of them (IRAS 06104+1524 and IRAS 06103+1523, respectively).  The IRAS point
sources both have kinematic distances of 3.5 kpc \citep{wouterloot89},
suggesting the subclusters are related.  The radio kinematic distances derived
to the two sources are the same, and the angular separation is comparable to
the size of either subclump; in addition, a slight overdensity of stars can be
seen in the $K$ image.  These factors suggest that, at the very least, these
clusters are related; they may differ in age, but they are likely to be part of
the same general star-formation event.  We see no difference between the two
apparent in either the CMD or the color-color diagram
(Figures~\ref{6104cmd}~and~\ref{6104cc}); if they do differ in age, it is
beyond the ability of our data to discern.  Due to this apparent association
and the small number of stars in each cluster, we analyze the two together as a
single cluster.  The CMD after statistical correction and adjusting to a
distance of 3.5 kpc \citep{wouterloot89} is shown in
Figure~\ref{6104-statcorr}.

Using the color-magnitude diagram based method of deriving an IMF as described
above, for a limiting extinction of $A_V = 25$, we find an IMF slope of $\Gamma
= -0.9 \pm 0.25$.  Using the KLF method (with no extinction correction, to
mimic the results of a study with only single-color photometry available) we
arrive at $\Gamma = -2.6 \pm 0.3$.  Even after allowing for the errors to be
larger than quoted due to uncertainty in the photometry and the conversion to
mass, these two slopes are inconsistent with each other, suggesting that
systematic effects in one or both methods dominate over the statistical errors.

\subsection{Sh 2-288}

\begin{figure}
\includegraphics[scale=.8]{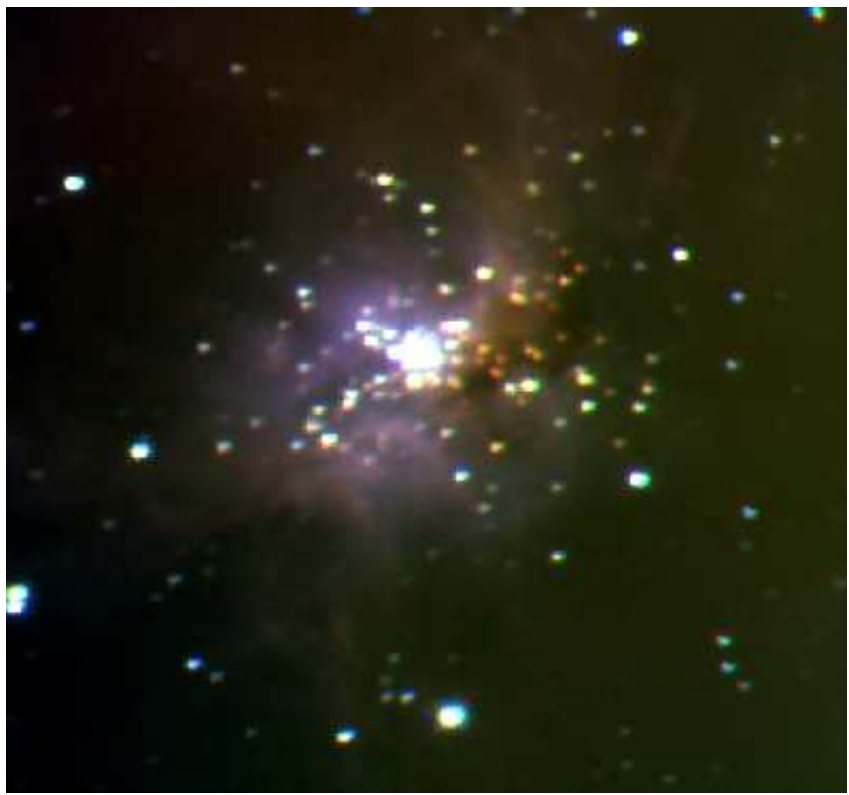}
\figcaption{\label{70839image} Color composite ($J$ = blue, $H$ = green, $K$ =
red) of the Sh 2-288 cluster region (North=up, East=left).   }
\end{figure}

The near-IR cluster image (Figure~\ref{70839image}) of the cluster near Sh2-288
shows a cluster with a dense core, crossed near the center by a dust lane.  The
center of the cluster is unresolved in our images, taken with a seeing of
$0.7\arcsec$.  This region was previously identified as an embedded cluster by
\citep{db01}.  In their catalog of outer-Galaxy HII regions, \citet{rudolph96}
quote widely disparate distances, with a radio kinematic distance of $7.2$ kpc
and a photometric distance of $3.0$ kpc \citep{brandblitz93}.  The kinematic
distance would make the sources we observe (several with $K < 12$) extremely
massive, it is thus far more likely that the photometric distance is correct,
and we have used the photometric distance for our analysis.

The 8~$\micron\ $ image from the MSX mission shows a peak coincident with the
near-IR nebulosity; there is not a significant amount of 8~$\micron\ $ emission
from regions dark in the NIR.  This suggests that there are not a significant
number of sources so deeply embedded that they cannot be seen in $K$ present in
this cluster.

\begin{figure}
\scalebox{0.75}{\includegraphics[angle=-90]{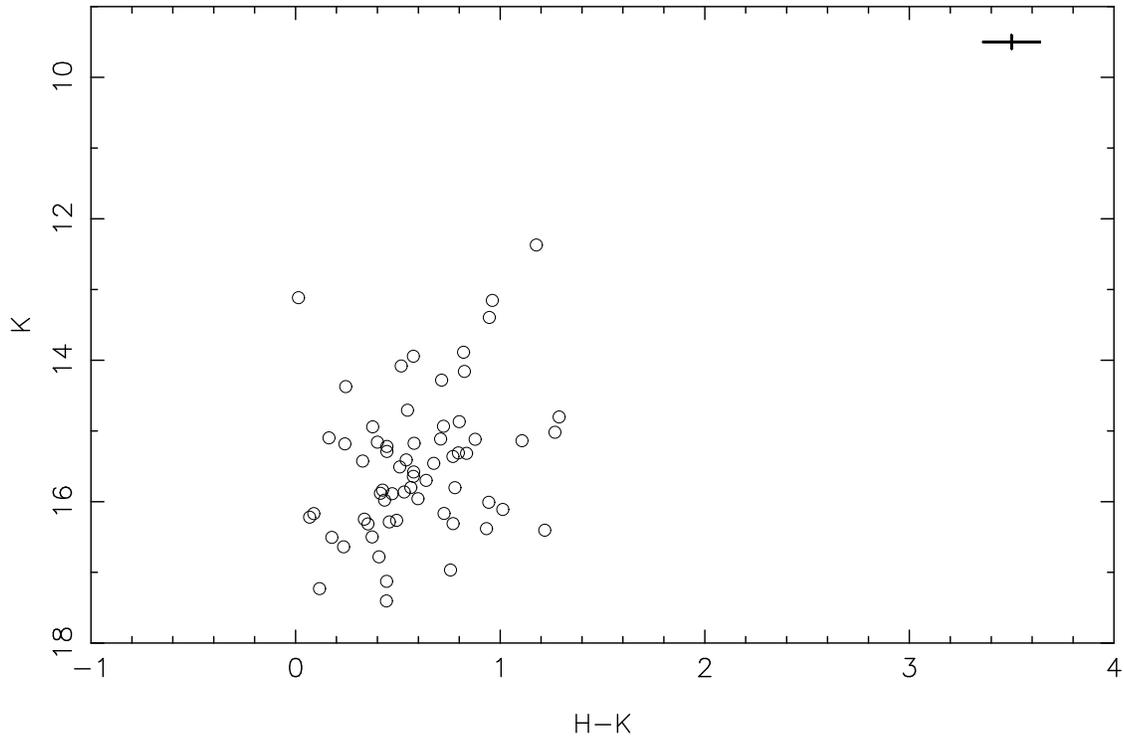}}
\figcaption{\label{70839cmd}  Statistically corrected $K$ vs. $H-K$ color-magnitude diagram for
the Sh2-288 cluster.  The bright central source, which we have identified as a
blend of two or more stars, has been removed as well.  Typical
errorbars are the size of the plot symbols. Overall
uncertainties including calibration (which include terms that will not affect
the relative position of the points) are indicated by the symbol in the upper right.}
\end{figure}

\begin{figure}
\scalebox{0.75}{\includegraphics[angle=-90]{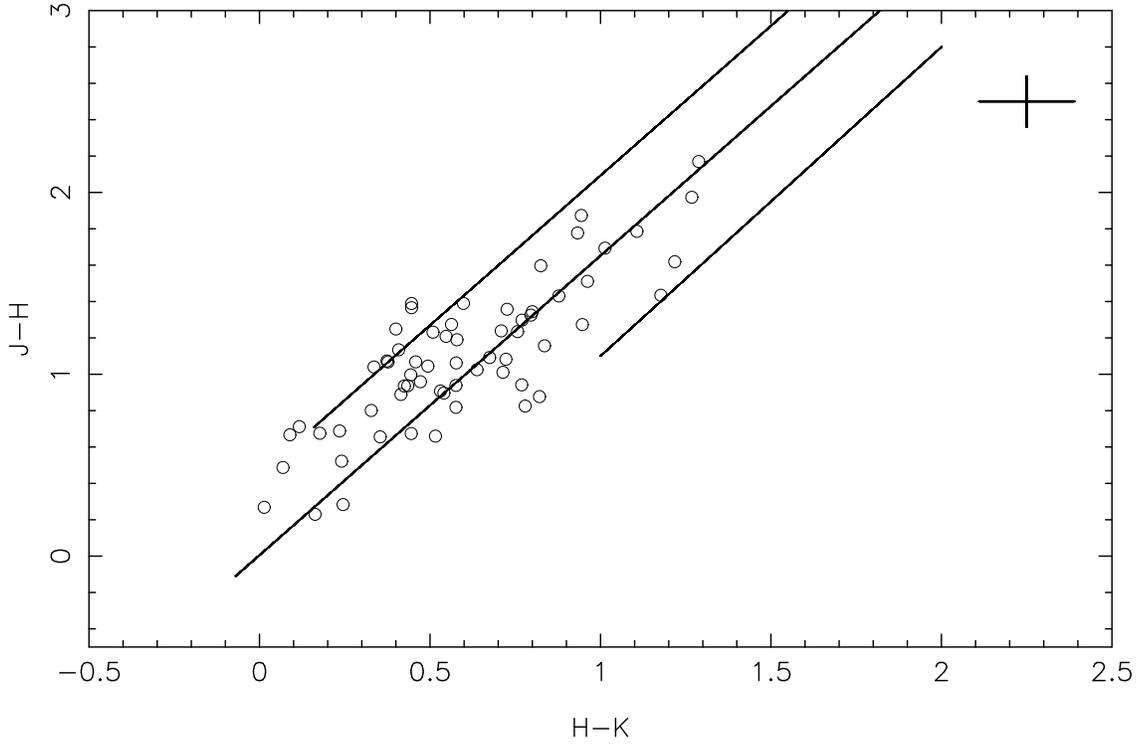}}  
\figcaption{\label{70839cc} Statistically corrected $J-H$ vs. $H-K$
color-color diagram or the Sh2-288 cluster.  The lines (parallel to the
reddening vector) delineate the possible location of reddened
main-sequence stars and of reddened T Tauri stars.  Typical
errorbars are the size of the plot symbols. Overall
uncertainties including calibration (which include terms that will not affect
the relative position of the points) are indicated by the symbol in the upper right.} 
\end{figure}

The $K$ versus $H-K$ color-magnitude diagram of the cluster near the HII region
Sh2-288 (Figure~\ref{70839cmd}) clearly shows the effects of variable
extinction; the stars separate into two groups, one nearly unextincted and one
with $\sim A_V = 5$.  We correct for field star contamination using the
region of the field outside the cluster region as described above.  The $J-H$
versus $H-K$ color-color diagram (Figure~\ref{70839cc}) shows few stars
separated from the main sequence locus by more than 2$\sigma$, suggesting that
most stars in this cluster are on the main sequence.  Extreme outliers in the
color-color diagram were inspected individually; in general, they lie in the
crowded central region of the cluster and most likely suffer from poor
photometry due to the different PSFs in $H$ and $K$ that resulted from
variations in seeing.  Such sources were excluded from analysis of the KLF and
the IMF.  Additionally, the brightest source in the cluster, which lies in the
most crowded central region and has a FWHM slightly broader than most sources
in the field, was excluded since it is quite likely to be a blend of multiple
sources.  We consider that the effects on the IMF are likely to be worse if a
blend is included than if any single star, even the most massive, is excluded.

Using the photometric distance of 3.0 kpc from \citep{rudolph96}, we
derive an IMF from the KLF of $\Gamma = -1.95 \pm 0.62$.  To better compare the
two methods of deriving the IMF, we included only those sources which were also
detected in $H$, so that the same dataset would be used for both the KLF and
CMD methods of deriving the IMF.    We individually de-reddened sources in the
$H-K$ CMD until they were on the main sequence, imposing an extinction limit as
before, and derive an IMF of $\Gamma = -1.62 \pm 0.65$.  Given the large
uncertainties, these results are entirely consistent.  The better agreement may
be because the extinction bias is less in the latter case.

\subsection{Comparison of Methods for IMF Determination}
\label{sec:compare}

A summary of the IMFs derived for the three clusters without significant
numbers of pre-main-sequence stars and the similar results from \citep{paper1}
is shown in Table~\ref{IMFsummary}.  In each case, there is a significant
difference between the IMF derived from the KLF and that derived from the CMD.
Does this reflect only uncertainties, or is one method in general more reliable
than the other?  A simple analysis would suggest that the CMD method is more
reliable, simply because it uses more information; the extinction clearly
varies across many embedded clusters (of those analyzed here, most notably Sh
2-288 and IRAS 06058+2158), and accounting for this should provide a more
robust estimate of the true IMF.  We cannot say for certain that this is the
case, however, without obtaining spectra for most of the stars in each cluster,
so that we can classify them spectroscopically and obtain individual masses.
We note that in each case, the IMF we derive from the CMD by individually
correcting the extinction for each object is flatter than that we derive from
the KLF by assuming a single extinction for the entire cluster.  This suggests
that more massive stars may preferentially lie in more heavily extincted
regions in embedded clusters.  Resolving this seeming discrepancy would require
spectra for a large number of cluster members, so that the IMF derived from
spectral classification of stars can be compared to that derived via the
different photometric methods.

\begin{figure}
\scalebox{0.75}{\includegraphics[angle=-90]{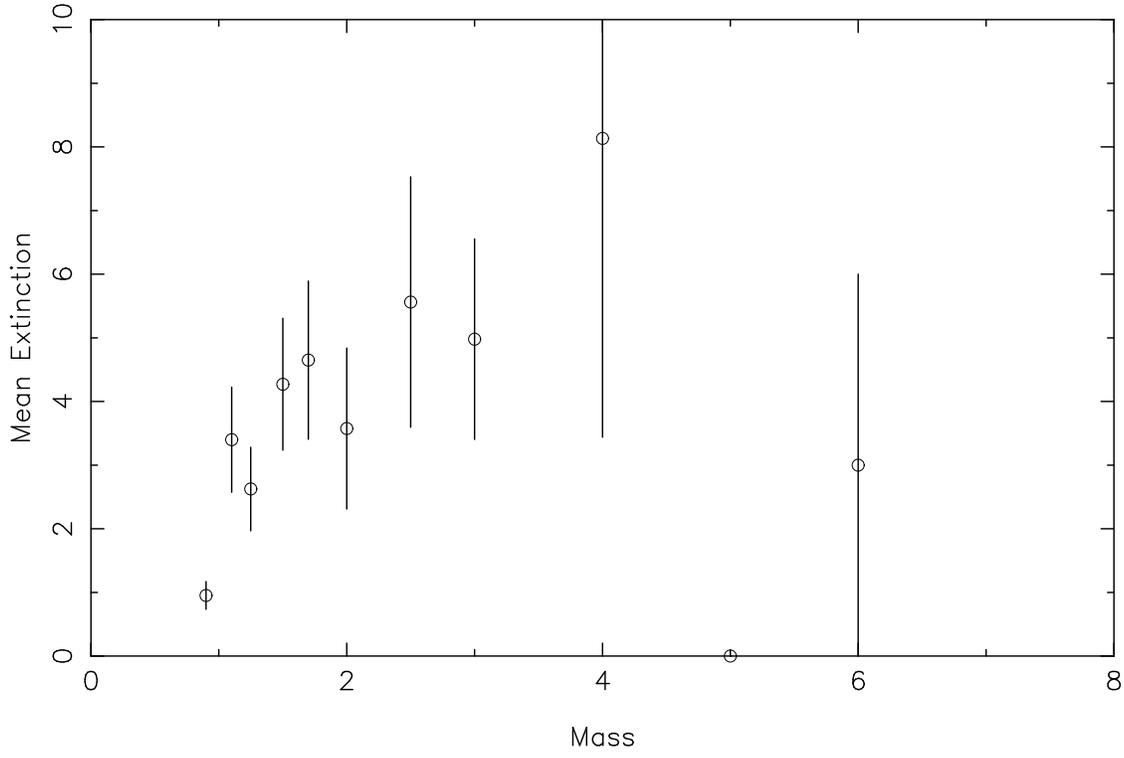}}
\figcaption{\label{fig:massext} Mean extinction derived from $H-K$ color 
for sources in each mass bin for the Sh2-217 cluster.  An extinction limit of 
$A_{V} = 10$ has been imposed; sources at higher derived extinction are not
included in the calculation.  More massive sources appear to preferentially lie
at higher extinctions, albeit with low significance.}
\end{figure}

We examine the relation between derived mass and extinction (with the
extinction limit imposed) for the Sh2-217 cluster, where we have the most data,
in Figure~\ref{fig:massext}.  Such a relation appears to be present, albeit at
low significance.  This effect is opposite in sign to what would be expected
from massive stars clearing their immediate environment more rapidly than their
lower-mass counterparts, but the stars we observe are mostly of intermediate
mass, rather than truly high mass, such that their winds are not as
significant; the effects of mass segregation, placing the more massive stars at
denser regions in the cluster, appear to dominate over the effects of clearing
in these clusters.  Since the clusters are numerically dominated by low-mass
stars, the average extinction will be mostly determined by the average for the
low-mass stars, changed slightly by the average for high-mass stars.  If more
massive stars are indeed preferentially found at higher extinction as our
results suggest, this would mean the majority of (low-mass) stars are
over-corrected for extinction by a small amount when using a single value,
while a few (high-mass) stars are under-corrected (thus lowering the derived
mass) by a large amount; thus, the effects of over-correcting and
under-correcting extinction do not fully cancel out, since the large
under-corrections would be more likely to move stars between mass bins than the
small over-corrections.  If there is no relation between mass and extinction,
we would expect these two results to cancel, since the average extinction for
low-mass stars would be the same as that for high-mass stars.

\section{Summary}

We present NIR images of four embedded clusters in the outer Galaxy.  In the
case of the cluster near IRAS 06058+2158 the number of stars with NIR excess
indicates a pre-main-sequence fraction between 60\% and 80\% and an age of less
than 3 Myr; the other three clusters show less nebular emission and fewer stars
with NIR excess indicating an older age.  We compute the IMF for the three
clusters dominated by main-sequence stars, in each case using both a KLF-based
method relying on a single extinction value for the cluster and using only $K$
band data and a CMD-based method where an individual extinction value is
calculated for each star.  We found a statistically significant difference
between the two values in two of the three cases, prompting us to examine IMF
values of embedded clusters from the literature to determine whether systematic
effects are at work.  We found a significant difference in the mean value of
the IMFs for embedded clusters derived from methods that handle extinction
individually compared with those that adopt a single value for the extinction.
Although a larger sample would help to make this claim more robust, since many
of the results come from a single study \citep{porras99} and methodological
details of that work could affect the results, we consider it to be significant
enough that IMFs obtained by different methods should not be compared in an
attempt to search for variations in the IMF from region to region.

Truly reliable IMFs for embedded clusters will most likely require spectra for
a large number of stars in the clusters; we are continuing to try to obtain
spectra for these sources to better characterize the massive star population
and the IMF of these clusters.

\clearpage

\acknowledgements

We thank Don McCarthy for use of and assistance with the PISCES instrument.  We
thank Phil Massey and Michael Meyer for comments that improved this paper.
This research has made use of the SIMBAD database, operated at CDS, Strasbourg,
France.  This publication makes use of data products from the Two Micron All
Sky Survey (2MASS), which is a joint project of the University of Massachusetts
and IPAC, funded by NASA and NSF.

\clearpage

\begin{deluxetable}{lllll}
\tablecaption{\label{IMFsummary}{IMFs for embedded stellar clusters}} 
\tablewidth{0pt}
\tablecolumns{5}
\tablehead{
\colhead{Cluster} &
\colhead{IMF: Common $A_V$} &
\colhead{IMF: Individual $A_V$} &
\colhead{Distance (kpc)} &
\colhead{Source}
}
\startdata
Sh2-288		&  $\Gamma=-1.95 \pm 0.82$	& $\Gamma=-1.62 \pm 0.5$ & $3.0$ &	This work\\
IRAS 06104+1524	&  $\Gamma=-2.49 \pm 0.3$	& $\Gamma=-1.38 \pm 0.6$ & $3.5$ &	This work\\
Sh2-217		&  $\Gamma=-2.7  \pm 0.25$	& $\Gamma=-1.61 \pm 0.2$ & $5.0$ &	This work\\
G305.3+0.2      &  $\Gamma=-1.5  \pm 0.3$       & $\Gamma=-0.98 \pm 0.2$ & $4.0$ &	Paper I\\
\enddata
\end{deluxetable}

\clearpage

\end{document}